\documentclass{aa}
\usepackage{graphics,latexsym,amssymb,times,psfig}

\def\gsim{ \lower .75ex \hbox{$\sim$} \llap{\raise .27ex \hbox{$>$}} }
\def\lsim{ \lower .75ex\hbox{$\sim$} \llap{\raise .27ex \hbox{$<$}} }

\begin{document}

\title{Low power BL Lacertae objects and the blazar sequence}

\subtitle{Clues on the particle acceleration process}

\author{Gabriele Ghisellini \inst{1}, Annalisa Celotti \inst{2}, Luigi
Costamante \inst{3,1} }

\offprints{G. Ghisellini; gabriele@merate.mi.astro.it}
\institute{Osservatorio Astronomico di Brera, via Bianchi 46, 
I--23807 Merate, Italy;
\and SISSA/ISAS, via Beirut 2-4, I--34014 Trieste, Italy;
\and Universit\`a Statale di Milano, via Celoria 16, I--20133 Milano, Italy}

\date{Received 2001}
 
\titlerunning{Low power BL Lacs}
\authorrunning{G. Ghisellini, A. Celotti \& L. Costamante}
   
\abstract{
The spectral properties of blazars seem to follow a phenomenological
sequence according to the source luminosity.
%We investigate the possible mechanisms responsible for
%such a trend in the spectral energy distributions.
By inferring the source physical parameters through (necessarily)
modeling the blazar spectra, we have previously proposed that the sequence
arises because the particles responsible for most of the emission
suffer increasing radiative losses as the luminosity increases.
Here we extend those results by considering the widest possible range
of blazar spectral properties.
We find a new important ingredient for shaping the spectra of the lowest 
power objects, namely the role of a finite timescale for the injection of 
relativistic particles. 
Only high energy particles radiatively cool 
in such timescale leading to a break in the particle distribution: 
particles with this break energy are those emitting most of the power,
and this gives raise to a link between blazar spectra and total energy density
inside the source, which controls the cooling timescale.
%, which controls the radiative cooling time. 
%, as also suggested by the rapid variability of BL Lac objects.
The emerging picture requires two phases for the particle acceleration:
a first pre--heating phase in which particles reach a
characteristic energy as the result of balancing heating and 
radiative cooling, and a more rapid acceleration phase which
further accelerate these particles to form a power law distribution.
While in agreement with standard shock theory,
this scenario also agrees with the idea that the luminosity of 
blazars is produced through internal shocks, which 
%i.e. collisions of different parts of the jet plasma moving at 
%different speeds, 
naturally lead to shocks lasting for a finite time.  }

\maketitle

\section{Introduction}

Blazars appear to come in different flavors, according to e.g. their strong
or weak (or even absent) broad emission lines,  their optical
polarization and the position of the peak of the synchrotron
component in their spectral energy distribution (SED), namely at low
(mm--IR) or high (UV and soft X--ray) frequencies (e.g. Padovani \&
Urry 2001).
The latter criterion has been indeed (proposed and) favored in the
last few years as possibly associated with more `physical' and
fundamental properties of the sources (Giommi \& Padovani 1994;
Fossati et al. 1998, hereafter F98; Ghisellini et al. 1998, hereafter G98).
In particular, within this scenario, F98 and G98
proposed that the different sub--classes of blazars form a spectral
and physical sequence in the position and intensity of the peaks, 
controlled primarily by one parameter, which
they identified with the bolometric observed luminosity.  
The sequence in the SED has been described by F98
phenomenologically (on the assumption that the
bolometric luminosity scales linearly with the radio one), while G98
modeled the SED of individual blazars adopting a homogeneous synchrotron 
plus inverse Compton model. 

In this way G98 deduced values of the physical quantities behind 
the sequence, i.e. translated the spectral trend (power vs SED)
into a trend between physical parameters, which could be consistently
interpreted as governed by the importance of radiative cooling.
In fact they found an inverse correlation between the energy of the 
particles emitting at the peaks of the SED, $\gamma_{\rm peak}m_{\rm e}c^2$, 
and the energy density $U$ (magnetic and radiation fields) as seen in the 
frame comoving with the emitting plasma.  
More specifically the correlation appeared to be well approximated by
$\gamma_{\rm peak} \propto U^{-0.6}$, directly implying that the radiative
cooling rate at $\gamma_{\rm peak}$ ($\propto \gamma^2_{\rm peak} U$) is
almost constant for all sources, suggesting a key role of the
radiative cooling process in shaping the SED.

Given the potential relevance of such result in the attempt to
understand the dissipation and particle acceleration mechanisms, we
decided to further explore the presence of such a trend over a larger
range of the parameter space.
In fact, in order to efficiently constrain the model parameters (in particular
from the high emission component) the modeling procedure in G98 was
applied only to those sources detected in the $\gamma$-ray band by EGRET
(the high energy instrument onboard the Compton Gamma Ray
Observatory), for which both the $\gamma$--ray spectral shape and
the cosmological distance were determined.

As the majority of the EGRET detected blazars comprises sources of
large power, Flat Spectrum Radio Quasars (FSRQ) and Low energy peak BL
Lacs (LBLs, in the definition of Giommi \& Padovani 1995), such
initial selection of the sources was biased against low power BL Lacs,
under--represented in G98, and thus against sources with high energy
peaks at very high frequencies (above the EGRET range).

In reality a few of those sources have been detected by
Cherenkov telescopes in the TeV band and the modeling of their broad
band energy distribution confirmed the phenomenological trend between
the SED and the source power (F98).  
Furthermore a number of low power BL Lacs have been recently observed
by $Beppo$SAX and detected in the X--ray band up to $\sim$ 100 keV.
Therefore there is now a reasonable number of comparatively weak
sources with sufficient data to constrain the shape of their SED and
thus explore the $\gamma_{\rm peak}$--$U$ correlation at high
$\gamma_{\rm peak}$ (low $U$).  

Such an extension becomes even more relevant since a change in
the correlation is indeed expected.  
In fact we already stress here that it is possible to immediately
predict a deviation from the $\gamma_{\rm peak} \propto U^{-0.6}$
behavior.  
Consider for illustration the case of the most extreme
sources, e.g. Mkn 501 in the flaring state and 1ES~1426+428 (Pian et
al. 1998; Costamante et al. 2001).  
They are both TeV detected sources (Catanese \& Weekes 1999 
and references therein; Horan 2000, Djannati--Atai,
priv. comm.) and also showed a
synchrotron component peaking above $\sim$ 100 keV.  
The peak at TeV energies implies that $\gamma_{\rm peak}$ exceeds
$10^5$--$10^6$: according to the correlation found by G98 this
would correspond to magnetic fields weaker than $\sim$ 0.05 G.  
However this implies a synchrotron peak frequency two--three orders of
magnitude smaller than observed. 
We therefore expect a change in the correlation at the high $\gamma_{\rm peak}$
end of the parameter space, and so plausibly we expect to find
an additional key ingredient, besides radiative cooling,
regulating the blazar SED. 

In this paper we thus explore such issue by extending the sample
of blazars, including several extreme high energy peakers,
as presented in Section 2.
We then consider at first the modeling of the SED of such objects with
the synchrotron inverse Compton model used by G98 and examine the
results in relation to the $\gamma_{\rm peak}$--$U$ correlation in Section 3.
The parameters inferred from the modeling imply that in the
extreme sources it is necessary to consider the effect of the duration
of the relativistic particle injection as this can be comparable or 
shorter than the relevant cooling timescale (see also Sikora et al. 2001,
Sikora \& Madejski 2001).
Therefore in Section 4 we re--analyze the whole sample of blazars with a
finite injection time model, discuss the findings and propose a physical 
interpretation for the $\gamma_{\rm peak}$--$U$ relation.
Summary and conclusions are reported in Section 5.

\section{The extreme BL Lac objects}

We consider ``extreme BL Lac objects'' the sources with a synchrotron
peak at energies exceeding 0.1 keV.
The frequency of the peak, $\nu_{\rm peak}$, can be thus directly
determined by broad band X--ray detectors, such as those onboard the
{\it Beppo}SAX satellite.

Candidate extreme BL Lacs were selected by Costamante et al. (2001) on
the basis of their SEDs and their broad band (radio/optical/X--rays)
spectral indices, which constitute good indicators of the location of
the synchrotron peak, as objects of different characteristics tend to
gather in different regions of the spectral index parameter spaces
(Stocke et al. 1991; Padovani \& Giommi 1995; F98).
For five out of the seven objects observed by {\it Beppo}SAX
(Costamante et al. 2001) $\nu_{\rm peak}$ turned out to be indeed in the
X--ray range and for one of them, 1426+428, $\nu_{\rm peak}$ was bound to
be at frequencies larger than 100 keV, as occurred in Mkn 501 during
the flaring state of 1997 (Pian et al. 1998) and in 1ES 2344+514
(Giommi, Padovani \& Perlman 2000). 
These three, so far unique, sources have all been detected at TeV
energies.
In addition to these, here we also reconsider Mkn 501 and 1ES
2344+514, together with Mkn 421, for modeling their SED observed
during flaring states (the SED in G98 corresponded only to the more
``quiescent'' states) and include in the sample PKS 2005--489 and 1ES
1101--232, since {\it Beppo}SAX observations of these sources revealed
a high peak synchrotron frequency (Wolter et al. 2000;
Tagliaferri et al. 2001).

In order to extend the range covered by the $\gamma$--$U$ relation we
also considered, at the other end of the blazar sequence, three very
powerful FSRQ recently found at high redshift (Fabian et al. 2001a,b;
Moran \& Helfand 1997 and references therein).
In fact, although they do not have a complete information on their
high energy peak (not being detected by EGRET), their hard X--ray
spectra constrain the emission model rather tightly.

In Table 1 we list the sources included in this paper (and not already
present in the sample of G98) and the references to the data used in
constructing the SEDs shown in Fig.~1a,b,c.

\begin{table*}
\begin{center}
\begin{tabular}{|llll|}
\hline
Source      &Other name &$z$  &Ref\\
\hline
0033$+$595  &1ES      &0.086$^a$ &NED, Co01, Pe96, SG99, Br97\\        
0120$+$340  &1ES      &0.272     &NED, Co01, BS94 \\     
0548$-$322  &1ES, PKS &0.069     &NED, Co01, GAM95, Pe96, WW90, Ta95, Ti94 \\          
1101$-$232  &1ES      &0.186     &NED, Wo00, SG99 \\        
1101$+$384  &Mkn 421  &0.031     &NED, Ma95, Ma96, Sc97  \\     
1114$+$203  &RGB J1117+202 &0.139$^b$ &NED, Co01, Br97 \\         
1218$+$304  &1ES      &0.130     &NED, Co01, GAM95, Pi93, Sa94, Fo98\\          
1426$+$428  &1ES      &0.129     &NED, Co01, GAM95, La96, Sa93, Sa97, Fi94, Ho00 \\        
1652$+$398  &Mkn 501  &0.034     &NED, Pi98  \\     
2005$-$489  &PKS      &0.071     &NED, Ta01   \\     
2344$+$514  &1ES      &0.044     &NED, Gi00, SG99, Pe96, Fi94 \\     
2356$-$309  &H 2356--309  &0.165 &NED, Co01, Be92, Fa94 \\          
\hline
0525$-$334  &PMN      &4.41     & Fa01a\\
1428$+$422  &B3       &4.715    & Fa01b\\
1508$+$571  &GB       &4.301    & MH97\\
\hline
\end{tabular}
\vskip 0.3 true cm
\caption{List of the sources studied in this paper. The line separates
the extreme BL Lacs (above) and the very powerful FSRQ (below). 
a) Tentative, Perlman, priv. comm. as reported in
Falomo \& Kotilanen 1999.
b) redshift reported in Bohringer et al. 2000.
Sources of data:
Be92: Bersanelli et al. 1992;
Br97: Brinkmann et al. 1997;
BS94: Brinkmann \& Siebert 1994;
Co01: Costamante et al. 2001;
F98:  Fossati et al. 1998;
Fa01a: Fabian et al. 2001a;
Fa01b: Fabian et al. 2001b;
Fa94: Falomo et al. 1994;
% Fe96: Fegan 1996{\bf NON LA TROVO???}
Fi94: Fichtel et al. 1994;
GAM95: Giommi, Ansari \& Micol 1991;
Gi00: Giommi, Padovani \& Perlman 2000;
Ho00: Horan 2000;
La96: Lamer, Brunner \& Staubert 1996;
Ma95: Macomb et al. 1995;
Ma96: Macomb et al. 1996;
MH97: Moran \& Helfand 1997;
Pe96: Perlman et al. 1996;  
Pi93: Pian \& Treves 1993;
Pi98: Pian et al. 1998 (and references therein);
Sa93: Sambruna et al. 1993;
Sa94: Sambruna et al. 1994;
Sa97: Sambruna et al. 1997;
Sc97: Schubnell 1997;
SG99: Stevens \& Gear 1999;
Ta95: Tashiro et al. 1995;
Ta01: Tagliaferri et al. 2001 (and references therein);
Wo00: Wolter et al. 2000;
WW90: Worral \& Wilkes 1990.
}
\end{center}
\end{table*}

\begin{figure*}
\psfig{figure=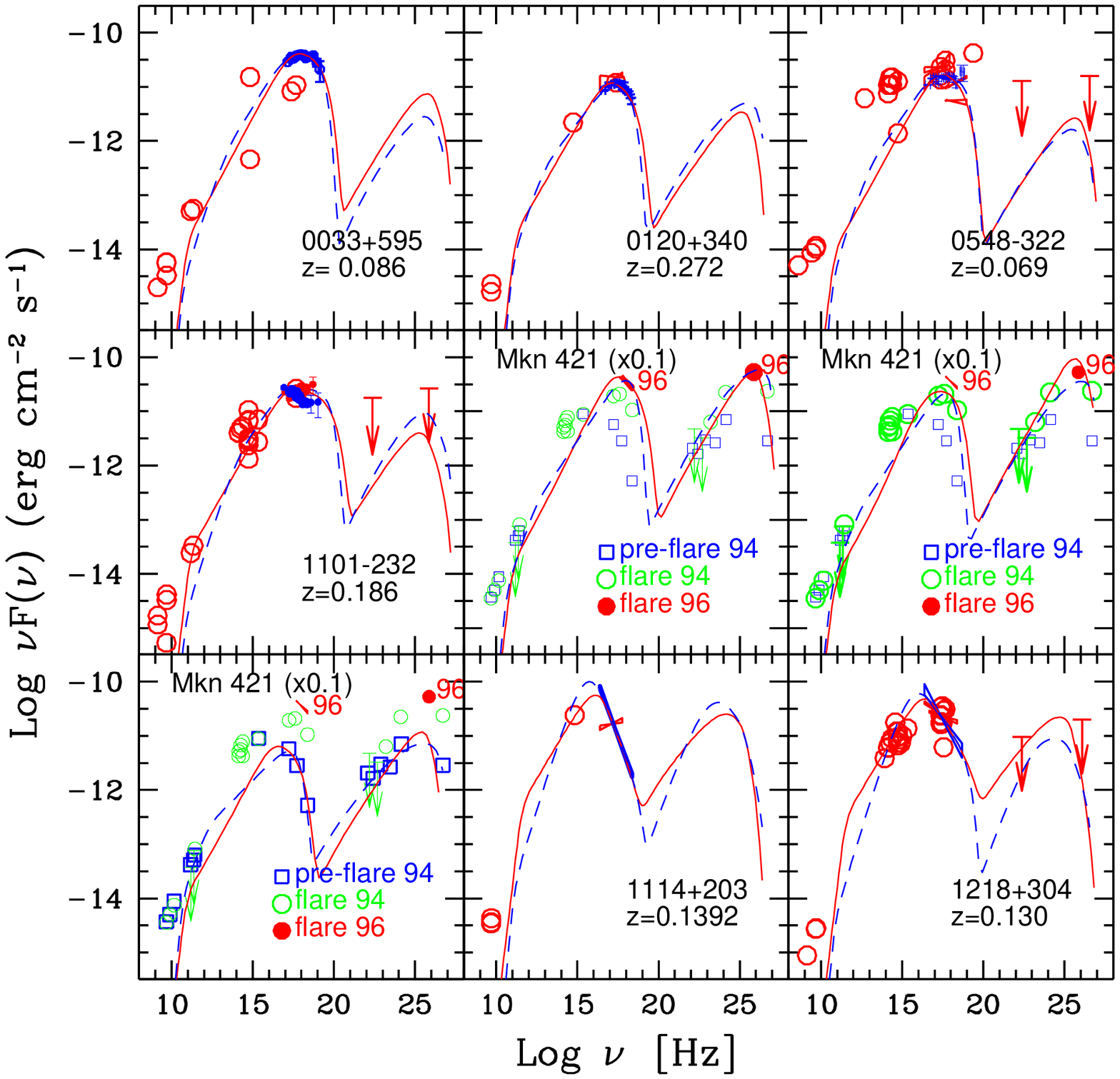,width=19cm}
\vskip -1 true cm
\caption{{a):} The SED of our extreme BL Lac objects. The solid and
dashed lines correspond to the steady--state and finite injection time
models, as discussed in Section 3 and 4, respectively.}
\end{figure*}

\setcounter{figure}{0}
\begin{figure*}
\psfig{figure=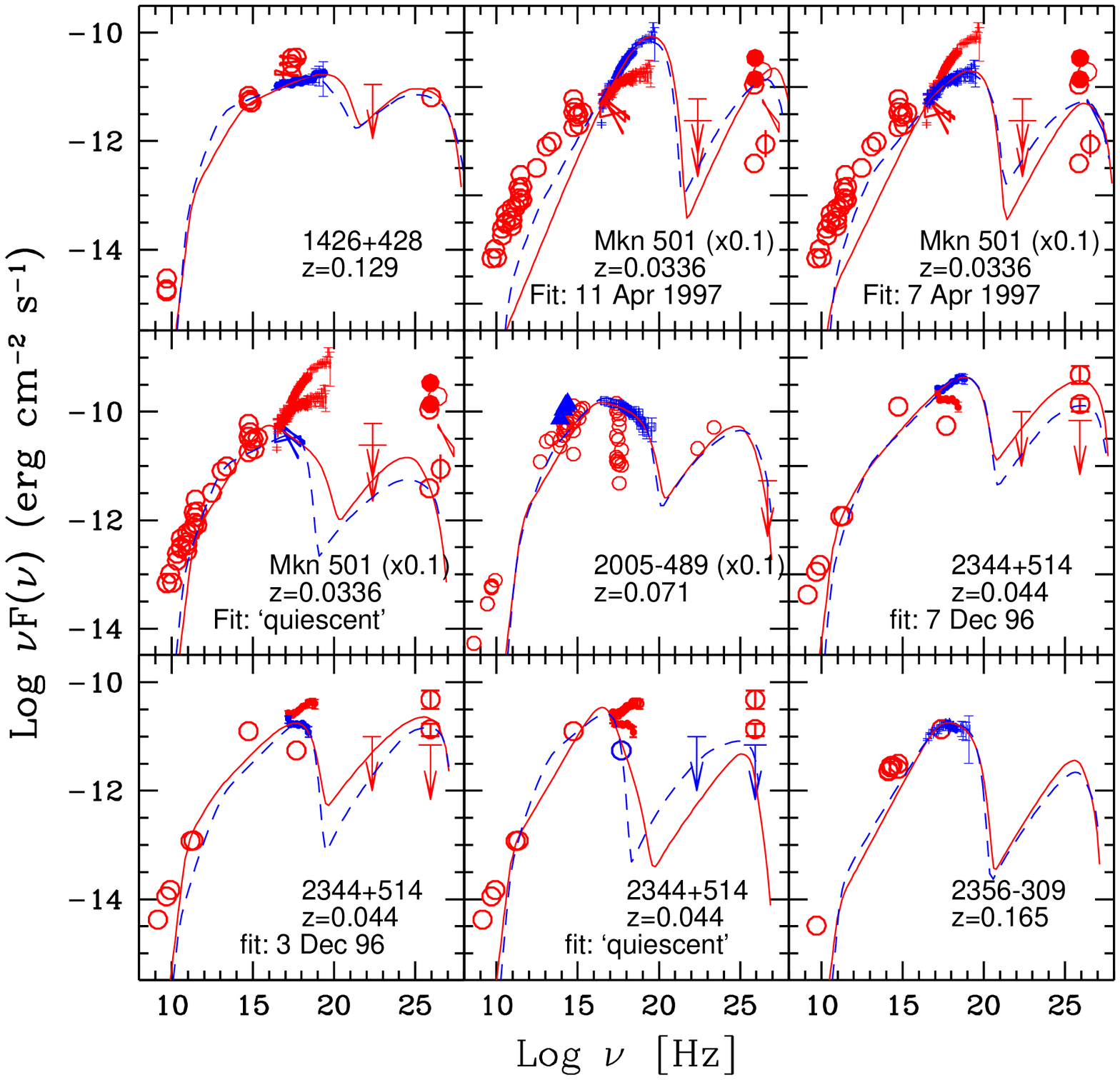,width=19cm}
\vskip -1 true cm
\caption{{b):} The SED of our extreme BL Lac objects. The solid and
dashed lines correspond to the steady--state and finite injection time
models, as discussed in Section 3 and 4, respectively.}
\end{figure*}

\setcounter{figure}{0}
\begin{figure*}
\psfig{figure=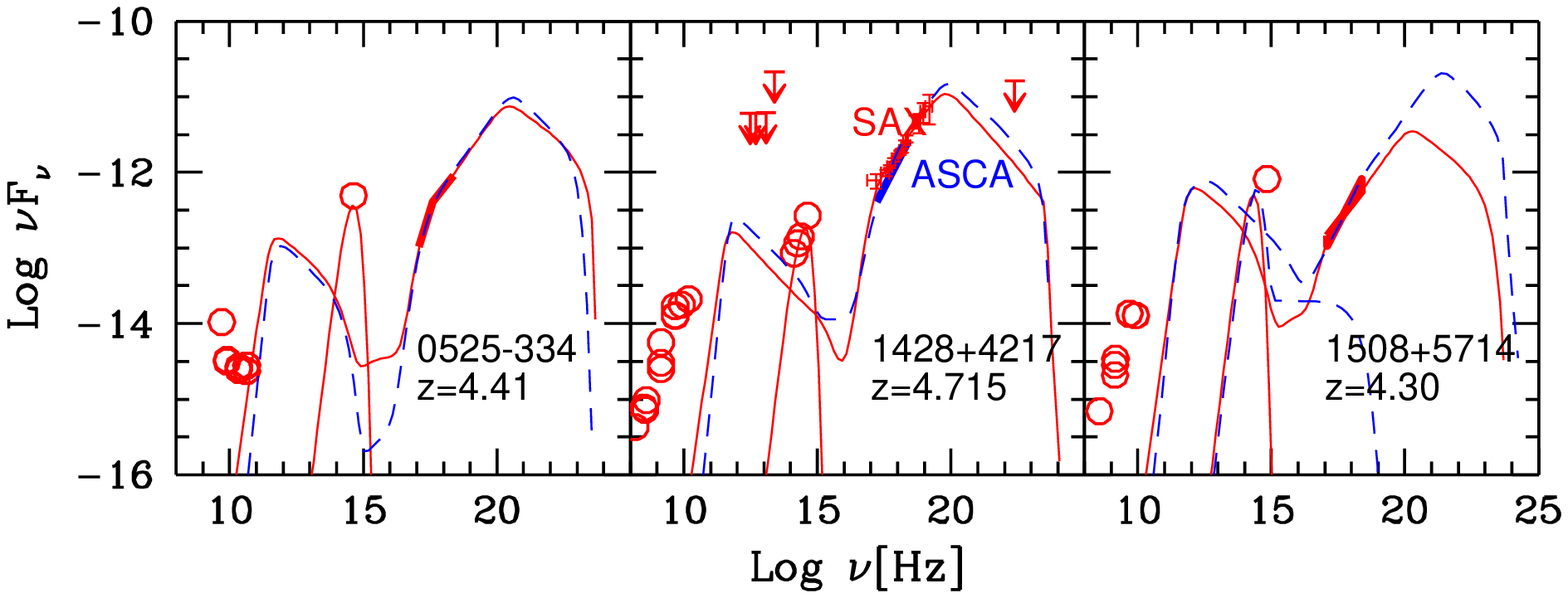,width=19cm}
\vskip -12 true cm
\caption{{c):} The SED of the 3 quasars at $z>4$ considered in this
paper.  The solid and dashed lines correspond to the steady--state and
finite injection time models, as discussed in Section 3 and 4,
respectively.}
\end{figure*}

\section{Steady--state synchrotron inverse Compton model}

\begin{table*}
\begin{center}
\begin{tabular}{|llllllllllll|}
\hline
Source      &$R$  &$L^\prime_{\rm inj}$ 
&$\gamma_{\rm min}$ &$\gamma_{\rm max}$ &$\gamma_{\rm peak}$  &$s$   &$B$   
&$\delta$ &$L_{\rm ext}$ &$R_{\rm ext}$  &Notes\\
            &cm &erg s$^{-1}$ & & & & &Gauss & &erg s$^{-1}$ &cm  &\\ 
\hline
0033$+$595 &5e15 &1.8e41 &9.0e4 &1.2e6  &9.0e4 &2.0 &0.73 &15  &--- &---  &   \\ %           
0120$+$340 &5e15 &5.6e41 &3.0e4 &3.0e5  &3.0e4 &2.2 &1.83 &15  &--- &---  &   \\ %    
0548$-$322 &5e15 &4.3e40 &8.0e4 &8.0e5  &8.0e4 &2.1 &0.51 &15  &--- &---  &    \\ %       
1101$-$232 &5e15 &6.5e41 &4.0e4 &1.5e6  &4.0e4 &2.1 &1.97 &15  &--- &---  &   \\ %       
1101$+$384 &5e15 &3.3e41 &1.0e5 &1.0e6  &1.0e5 &2.5 &0.34 &15  &--- &---  &1996 high state   \\
1101$+$384 &5e15 &1.1e42 &6.0e4 &8.0e5  &6.0e4 &2.0 &0.22 &11  &--- &---  &1994 flare (G98 fit) \\ 
1101$+$384 &5e15 &2.2e41 &4.0e4 &4.0e5  &4.0e4 &2.3 &0.27 &11  &--- &---  &1994 pre--flare   \\ 
1114$+$203 &5e15 &7.4e41 &1.2e4 &2.0e5  &1.2e4 &3.6 &2.11 &15  &--- &---  &   \\ %      
1218$+$304 &5e15 &6.5e41 &1.0e4 &4.0e5  &1.0e4 &3.0 &3.11 &15  &--- &---  &   \\ %       
1426$+$428 &5e15 &1.1e41 &6.0e3 &7.0e6  &1.0e6 &1.9 &0.52 &20  &--- &---  &   \\ %      
1652$+$398 &5e15 &1.7e41 &1.0e6 &9.0e6  &2.5e6 &1.4 &0.10 &20  &--- &---  &11 Apr  1997 \\
1652$+$398 &5e15 &4.6e40 &7.0e4 &5.0e6  &1.4e6 &1.4 &0.17 &20  &--- &---  &07 Apr 1997 \\ 
1652$+$398 &5e15 &3.7e41 &1.0e4 &8.0e5  &1.0e4 &2.8 &1.11 &10  &--- &---  &quiescent, G98 fit \\
2005$-$489 &5e15 &6.5e41 &1.0e4 &7.0e5  &1.0e4 &2.2 &2.20 &15  &--- &---  &   \\       
2344$+$514 &6e15 &8.9e40 &1.0e3 &4.0e6  &1.3e6 &1.5 &0.14 &15  &--- &---  &07 Dec 1996 high state SAX  \\ 
2344$+$514 &6e15 &4.4e40 &1.0e3 &1.0e6  &3.0e5 &1.5 &0.14 &15  &--- &---  &03 Dec 1996 low state SAX  \\ 
% 2344$+$514 &6e15 &3.3e40 &1.5e4 &5.0e5  &1.5e4 &2.6 &0.37 &15  &--- &--- & \\ % quiescent state (new fit)
2344$+$514 &8e15 &4.1e40 &4.0e4 &7.0e5  &4.0e4 &3.7 &0.47 &14  &--- &---  &quiescent, G98 fit \\  
2356$-$309 &5e15 &3.1e41 &7.0e4 &1.2e6  &7.0e4 &2.0 &0.97 &15  &--- &---  &   \\         
\hline
0525$-$334 &2e16 &1.4e44 &45    &2.0e3  &45    &2.5 &5.13 &15  &5.0e45  &8.0e17  &\\
1428$+$422 &2e16 &2.2e44 &25    &5.0e3  &25    &2.8 &9.12 &15  &3.0e45  &4.0e17  & \\
1508$+$571 &2e16 &7.4e43 &50    &2.0e3  &50    &2.5 &11.8 &15  &6.6e45  &1.2e18  & \\
\hline
\end{tabular}
\vskip 0.3 true cm
\caption{
Parameters used to model the SED of the extreme BL Lacs
(above the line) and powerful FSRQ (below the line) considered in this
work, according to the steady--state synchrotron inverse Compton
model as discussed in Section 3. $L_{\rm ext}$ and $R_{\rm ext}$ correspond 
to the luminosity and extension of the external 
photon source (assumed to originate in a Broad Line Region).}
\end{center}
\end{table*}

The SED of the considered sources have been modeled as in G98.
We briefly remind the main assumptions of this model: in a
spherical source of radius $R$, embedded in a tangled and homogeneous
magnetic field $B$, a distribution of relativistic electrons is
continuously injected at the rate $Q(\gamma)\propto \gamma^{-s}$
[cm$^{-3}$ s$^{-1}$] between $\gamma_{\rm min}$ and $\gamma_{\rm max}$, with a
corresponding injected power $L_{\rm inj}^\prime$ as measured in the comoving
frame.
Note that $\gamma_{\rm min}$ is the minimum Lorentz factor of the 
injected electrons, and should not be confused with the minimum 
Lorentz factor of the emitting particle distribution, which
is here assumed to extend down to $\gamma\sim 1$.
The emitting particle distribution and corresponding spectrum
are derived by solving the
steady--state continuity equation, assuming synchrotron and inverse
Compton radiative losses and possible pair production through
photon--photon collisions.
The seed photons for the inverse Compton process are both synchrotron
and external photons, the latter ones assumed to be distributed as a
(diluted) blackbody with peak frequency $\nu_o\sim 10^{16}$ Hz (as
observed in the comoving frame).
Externally produced radiation has been considered for the three
high redshift quasars, while for all the extreme BL Lacs studied here 
the spectrum can be fitted with a pure synchrotron self--Compton
(SSC) model, and we can therefore consistently neglect the presence of
externally produced radiation.

For sources with $\gamma$--ray information we have enough data 
to completely constrain the SSC model. 
For the remaining objects we have to assume from 
the variability timescale $t_{var}$ the parameters $R$ 
and the bulk Lorentz factor $\delta$.
We decided to adopt the same values of $\delta$ (=15) and 
$R$ ($=5\times 10^{15}$ cm) for all sources with no $\gamma$--ray data.
Such values are in agreement with what we inferred for Mkn 501
and 1ES 1426+428 and with values reported by other authors in the
literature (e.g. Mastichiadis \& Kirk 1997; 
Tavecchio, Maraschi \& Ghisellini 1998; Kataoka et al. 1999;
Kino, Takahara \& Kusunose, 2001).
In the following we examine the dependence of the results on the 
assumption on $\delta$
\footnote{The Lorentz factor of the bulk motion of the plasma is taken
to be equal to the Doppler factor (i.e. $\Gamma\sim \delta$),
i.e. corresponding to a viewing angle $\theta\sim 1/\Gamma$.}.
For the three high redshift quasars we have fixed the
value of $\delta$ (=15, as for the other BL Lacs) and assumed a 
somewhat larger size of the emitting region ($R=2\times 10^{16}$ cm).

Note that the applied model is aimed at reproducing the spectrum
originating in a limited part of the jet, thought to be responsible
for most of the emission.
This region is necessarily compact, since it must satisfy the
constraints from the fast variability shown by blazars especially
at high frequencies.
Therefore the radio emission from such compact regions is strongly
self--absorbed: the model cannot thus account for the observed radio
flux.

In Table~2 we list all the input 
parameters\footnote{No statistical significance of the modeling is
considered here. As discussed in G98 we are interested in reproducing
the average spectrum of a statistically significant number of
objects.}
used for the ``fits'' shown in Fig.~1a,b,c as solid lines, 
and we also report the value of the derived parameter 
$\gamma_{\rm peak}$. 

In Fig.~2 we plot $\gamma_{\rm peak}$ as a function of the (comoving) 
energy density $U=U_{\rm r}+U_B$ (radiative plus magnetic) for these 
sources and those of G98. 
As in G98, the plotted radiation energy density $U_{\rm r}$ 
is not the total one, but only that part available for scattering in 
the Thomson regime for electrons of Lorentz factor $\gamma{\rm peak}$.
Fig.~2 clearly shows that $\gamma_{\rm peak}$ and $U$ 
are still significantly correlated also for the more extreme objects, 
but, as expected, with a different functional dependence of 
$\gamma_{\rm peak}$ vs $U$: for
values of $\gamma_{\rm peak}$ greater than $\sim 1000$ (corresponding
to $U\lsim 1$ erg cm$^{-3}$), the new branch is approximately described by
$\gamma_{\rm peak}\propto U^{-1}$.  
In Table 4 we give the statistical
results of a linear correlation between $\log \gamma_{\rm peak}$
and $\log U$.

Let us consider the robustness of such finding with respect to our
assumption on the Doppler factors.  
As mentioned above for Mkn 501 and 1ES 1426+428 the model parameters
are completely determined by the spectral and variability information,
while for the other sources we had to assume a value for the beaming
factor $\delta$ (=15).
Using for all of the new sources a lower (higher) value of $\delta$ would
translate only into a shift of the $\gamma_{\rm peak}\propto U^{-1}$ branch
towards higher (lower) $U$, not affecting its slope and thus its
presence.\footnote{e.g. $\delta=10$ corresponds to a
change of $U$ by a factor $\sim 3$}
Clearly, a spread in the values of $\delta$ would result in a scatter
of the points around the correlation.
Even in the fine tuned assumption of a systematic increase of $\delta$
-- say between 10 and 20 -- along the branch towards high
$\gamma_{\rm peak}$, the slope would not change enough to recover the
$\gamma_{\rm peak} \propto U^{-0.6}$ correlation.
We conclude that the presence of the new steeper trend is robust. 

On the other hand, we stress that the three powerful objects have
been included in the sample in order to better test the presence of
the flatter $\gamma_{\rm peak} \propto U^{-0.6}$ branch. Indeed as
shown for example in the case of 1428+4217, the extension of the hard
X-ray spectrum clearly constrains a minimum value of $\gamma_{\rm
peak}$.

\subsection{Interpretation}

The existence of such good correlations between $\gamma_{\rm peak}$ and $U$
intriguingly suggests that these are the result of robust physical
process(es). 
While, as already mentioned, the high $U$ branch corresponds to a
constant (with respect to all other parameters) cooling rate at
$\gamma_{\rm peak}$, the new steeper branch $\gamma_{\rm peak} \propto U^{-1}$
is equivalent to an approximately constant radiative cooling time
$t_{\rm cool}(\gamma_{\rm peak})$.
This appears to indicate that $\gamma_{\rm peak}$ might be determined by the 
cooling timescale approaching another relevant timescale of the system.
Plausibly the most characteristic one would be the light crossing time
associated to the dimension of the system.
And indeed we find also quantitative indication from the parameters of the
model that $t_{\rm cool}(\gamma_{\rm peak}) \sim R/c$.
Processes which can operate on such a timescale include adiabatic losses
and/or particle escape (for expansion and escape velocities of order
$\sim c$, see Kino, Takahara \& Kusunose, 2001) or a particle injection 
(e.g. in a relativistic shock) lasting for $R/c$.

However the similarity of these two timescales also puts in evidence the
inconsistency of the assumed model.
More precisely if the typical cooling timescale for the particle
radiating the bulk of the emission is of the order of the light
crossing time, the assumption of steady state for deriving the
particle distribution cannot be satisfied.

In such a situation therefore we have to consider a different
scenario.
In this respect two observational facts should be taken 
into account for the modeling of the emission. 
The variability characterizing blazars in general, and low power
BL Lacs in particular, indicates that the deposition of energy is not
continuous, but rather suggests a finite time of injection for each
typical flare.
Furthermore the symmetry of the raise and decay of flux in the light
curves during flaring episodes, together with the absence of a plateau
at constant flux, also indicates that the injection timescale 
should not exceed $R/c$ (Chiaberge \& Ghisellini 1999).
For these reasons we also explore here the situation where the injection
lasts for a finite timescale, of order of the light crossing time.

A physical scenario where flares naturally occur and the
injection timescale can be of order of the dynamical one is that of
internal shocks (e.g. Piran 1999; Ghisellini 1999a; Spada
et al. 2001), in which the dissipation takes place during the
collision of two shells of fluid moving at different speeds. Let us then 
adopt such scenario. 

\begin{figure}
\vskip -0.5 true cm
\psfig{figure=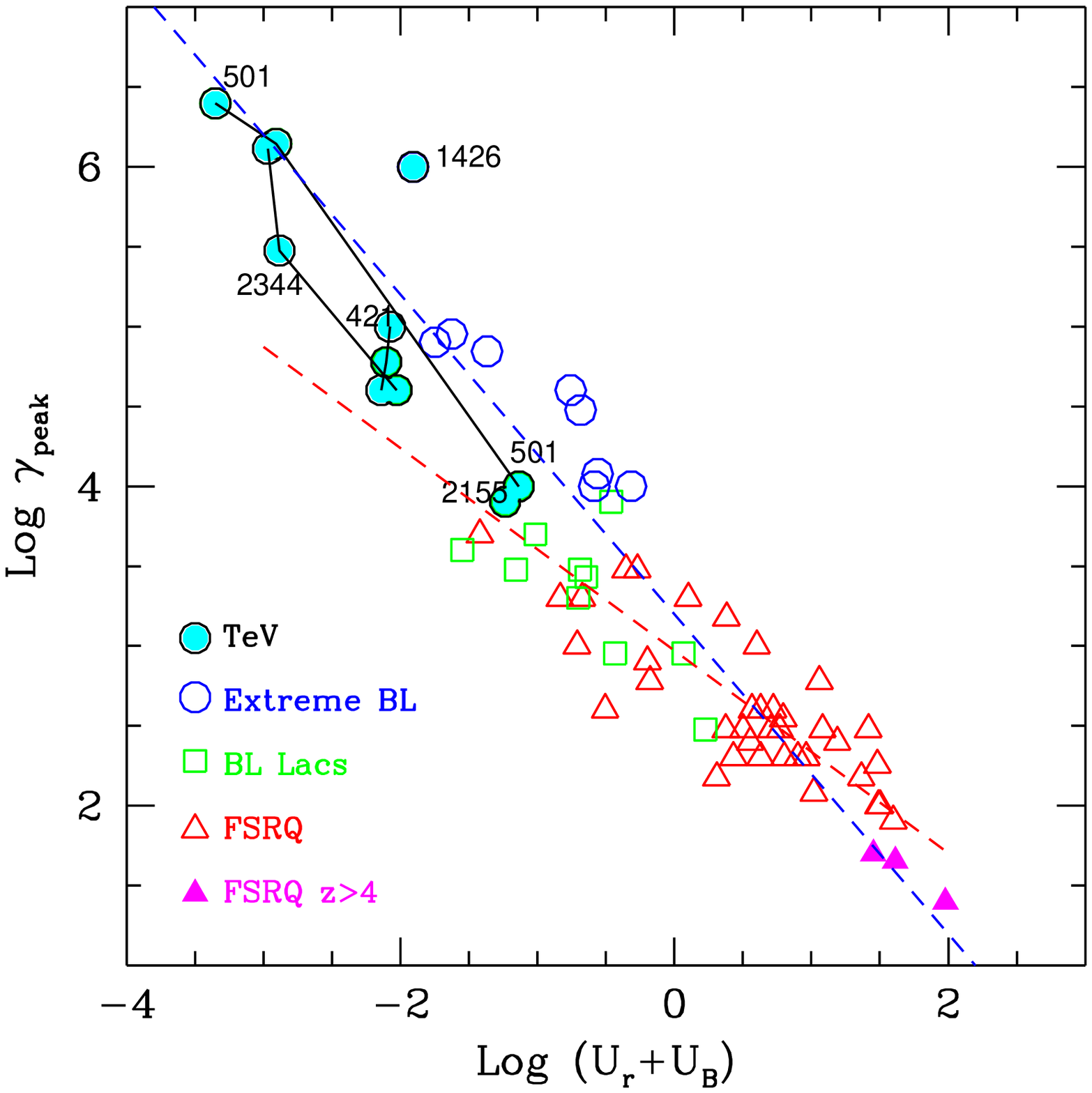,width=9.3cm}
\vskip -0.5 true cm
\caption{The Lorentz factor of the electrons emitting at the peaks of
the SED, $\gamma_{\rm peak}$, as a function of the comoving energy density
(radiative plus magnetic), according to the steady--state synchrotron
self--Compton model, as discussed in Section 3.  The points connected by
a line correspond to the quiescent and flaring states of the same source, 
namely Mkn 501, Mkn 421 and 1ES 2344+514, as labeled.
The dashed lines corresponds to the linear correlations found by G98 and
the one found in this paper considering only BL Lac objects
(see Table 4.}
\end{figure}

\section{Finite injection time model} 

\begin{table*}
\begin{center}
\begin{tabular}{|lllllllllllll|}
\hline
Source      &$R$  &$L^\prime_{\rm inj}$ &  $\gamma_{\rm min}$ &$\gamma_{\rm max}$ 
&$\gamma_{\rm peak}$  &$s$  &$B$ & $\delta$ &$L_{BLR}$ &$R_{BLR}$ &Notes\\
            &cm   &erg s$^{-1}$         &      & & & & Gauss &  & erg s$^{-1}$ &cm  & \\ 
\hline
%          R       L'      g1    g2     gpeak delta B  s     \\      gc  
0033$+$595 &9.0e15 &1.0e41 &1000   &6.0e5 &5.3e4 &2.01&0.8 &18.2  &--- &--- &   \\ %             
0120$+$340 &1.0e16 &4.4e41 &700    &2.5e5 &7.0e4 &2.2 &0.6 &18.6  &--- &---  &     \\ % 70406.80 
0548$-$322 &8.0e15 &7.5e40 &700    &5.0e5 &5.0e4 &2.1 &0.8 &14.1   &--- &--- &     \\ % 50271.88  
1101$-$232 &7.0e15 &5.3e41 &2000   &1.0e6 &5.3e4 &2.1 &0.9 &19.9   &--- &--- &     \\ % 52908.29 
1101$+$384 &1.5e16 &3.3e42 &1000   &7.0e5 &7.0e5 &2.1 &0.1 &14.3   &--- &--- &1996 flare \\ %  1359010. 
1101$+$384 &1.7e16 &1.4e42 &800    &5.5e5 &5.5e5 &2.1 &0.08 &14.3 &--- &---  &flare \\ % 1842846 
1101$+$384 &1.7e16 &1.5e41 &800    &3.8e5 &3.8e5 &2.4 &0.08 &14.3  &--- &---  &Pre--Flare \\ %1904686 
1114$+$203 &8.0e15 &4.5e41 &6000   &2.0e5 &6.0e3 &3.6 & 1.5 & 18.8 &--- &---  &  \\ %1.3e4       
1218$+$304 &1.0e16 &2.0e41 &6000   &3.0e5 &1.4e4 &2.9 & 1.5 & 19.3 &--- &---  &  \\ % 14185.67 
1426$+$428 &1.3e16 &7.0e40 &1000   &6.0e6 &7.2e5 &2.7 &0.18 &25.1 &--- &---  &     \\  % 718176.4 
1652$+$398 &7.0e15 &7.0e42 &500    &4.0e6 &5.5e5 &1.9 &0.28& 16.3 &--- &---  &11 Apr 1997 \\ % 552539
1652$+$398 &7.0e15 &6.0e41 &200    &4.0e6 &4.5e5 &2.26&0.3 & 16.3 &--- &---  &07 Apr 1997 \\ % 447785.3 
1652$+$398 &1.8e16 &2.5e40 &1000   &3.0e5 &3.0e5 &2.7 &0.24 &16.3 &--- &---  &quiescent  \\   % 303878
2005$-$489 &1.0e16 &8.0e41 &600    &5.0e5 &1.3e4 &2.3 &1.5 &15.5   &--- &--- &          \\ % 11739.36  
2344$+$514 &1.2e16 &6.0e41 &2000   &2.4e6 &1.6e6 &2.4 &0.11 &14.2 &--- &---  &7 Dec 1996 high state   \\ %1606568.     
2344$+$514 &1.2e16 &1.2e41 &2000 &8.0e5 &8.0e5 &2.4&0.09& 14.2  &--- &---  &3 Dec 1996 low state    \\ %2958959.
2344$+$514 &1.e16  &5.0e40 &1000 &1.0e5 &1.0e5 &2.5&0.4 &14.2  &--- &---  &quiescent        \\  % 186510.2  
2356$-$309 &7.0e15 &2.0e41 &100  &7.0e5 &6.0e4 &2.01& 0.9 &18.8 &--- &---  &          \\ % 59652.09   
\hline
0525$-$334 &2.0e16 &6.0e43  &60   &1.0e3 &60.0  &2.7 &10 &19.0 &1.1e46  &2.0e17 &   \\
1428$+$422 &2.0e16 &5.0e44  &33   &2.0e3 &33.0  &2.7 &11 &12.6&8.0e45  &2.5e17 &    \\
1508$+$571 &3.0e16 &6.0e44  &200  &8.0e3 &200   &2.7 &14 &12.7&1.2e45  &4.0e17 &   \\ 
\hline
\end{tabular}
\vskip 0.3 true cm
\caption{Parameters used to model the SED of the extreme BL Lacs and
powerful FSRQ (above and below the horizontal line, respectively)
considered in this paper, according to the finite injection time
synchrotron inverse Compton model as discussed in Section 4.
$L_{BLR}$ and $R_{BLR}$ correspond to the equivalent luminosity and extension 
of the external photon energy density 
(assumed to originate in a Broad Line Region).} 
\end{center}
\end{table*}

\begin{figure}
\vskip -0.8 true cm
\psfig{figure=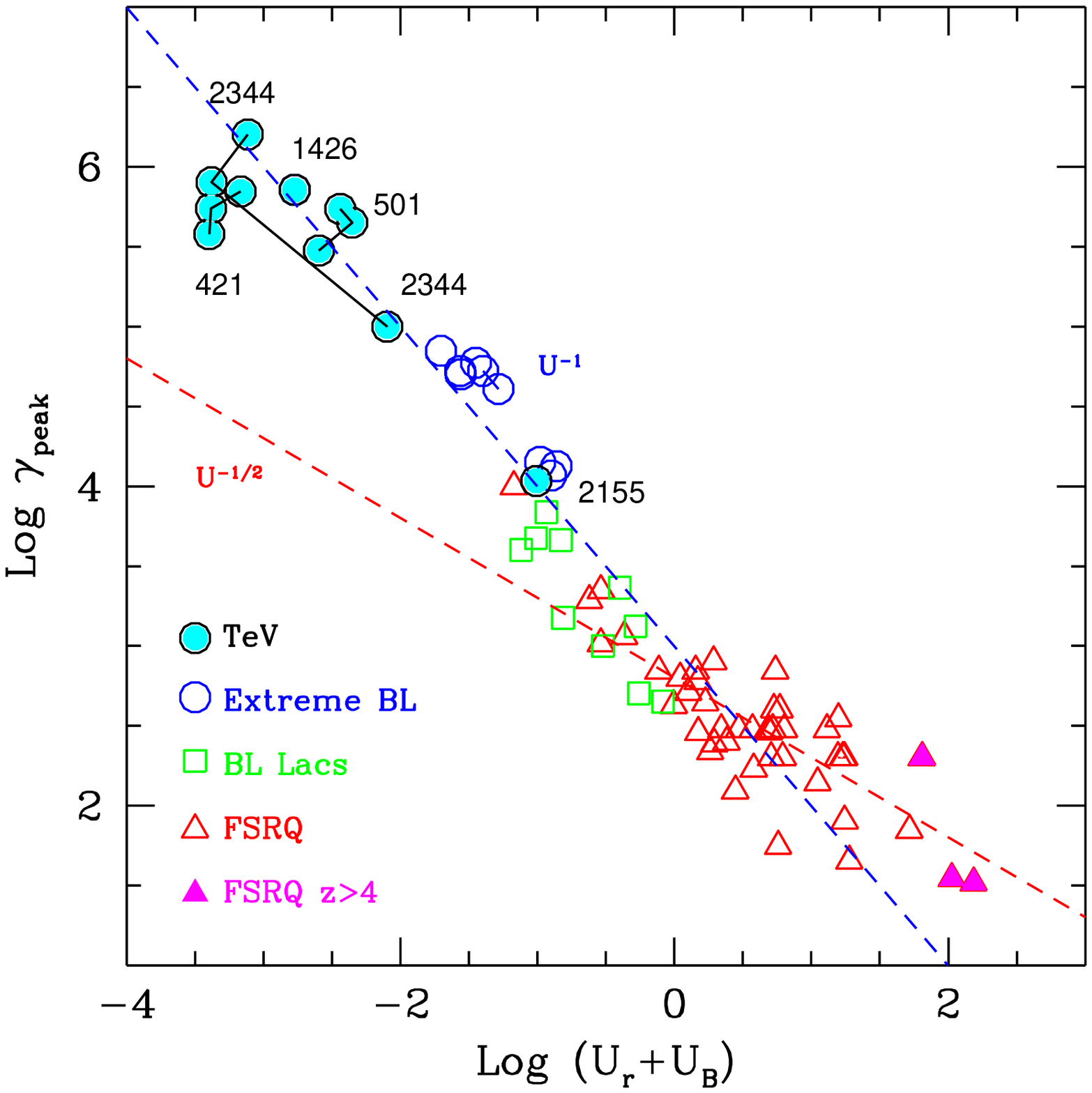,width=9.3cm}
\caption{The Lorentz factor of the electrons emitting at the peaks of
the SED, $\gamma_{\rm peak}$, as a function of the comoving energy density
(radiative plus magnetic), according to the finite injection time
synchrotron inverse Compton model as discussed in Section 4.  The
points connected by a line correspond to the quiescent and flaring
state of the same source, namely Mkn 501, Mkn 421, 1ES 2344+514, as
labeled.
The dashed lines correspond to $\gamma_{\rm peak} \propto U^{-1/2}$ and
$\propto U^{-1}$ (they are not best fits).}
\end{figure}

\begin{table*}
% \begin{center}
\begin{tabular}{llllll}
\hline
 Model      &$m$ &$q$ &$N$ &$r$ &$P$  \\
\hline
Steady st. (BL)   &$-$0.958$\pm$0.11  &3.10$\pm$0.18 &30  &$-$0.86  &7.3e-8  \\
Fin. inj. (BL)    &$-$0.975$\pm$0.06  &2.94$\pm$0.12 &30  &$-$0.95  &7.5e-8  \\
Steady st. (FSRQ) &$-$0.565$\pm$0.06  &2.88$\pm$0.06 &44  &$-$0.84  &4.4e-8  \\
Fin. inj. (FSRQ)  &$-$0.544$\pm$0.06  &2.79$\pm$0.05 &44  &$-$0.82  &1.5e-8  \\
\hline
\end{tabular}
\vskip 0.3 true cm
\caption{Results of linear correlations between $\log \gamma_{\rm peak}$ 
and the total energy density  $\log U$ (magnetic plus radiative) 
in the comoving frame. 
We have fitted separately all BL Lac objects, including the extreme
sources studied in this paper, and all FSRQ, including the three 
high redshift quasars studied here.
The correlation is in the form 
$\log \gamma_{\rm peak} = m \log U +q$. 
$P$ is the probability of a random distribution and $r$ the correlation
coefficient.}
\end{table*}

As in the steady state model just considered, the magnetic field is
assumed to be homogeneous and tangled throughout the region.
Note however that in the internal shock scenario the geometry of the
emitting volume is a cylinder with thickness $\Delta R'\sim R/\Gamma$
(in the comoving frame) and radius $R$ corresponding to the cross
section of the jet. 
In the following we thus adopt such geometry and 
again assume $\Gamma\sim\delta$.
%specify both the bulk Lorentz factor and the viewing angle.

In the hypothesis of a finite injection time $t_{\rm inj}$, which we
assume lasts for $t_{\rm inj}=\Delta R'/c$, clearly the continuity
equation governing the particle distribution has no steady state
solution (see also Sikora et al. 2001).
However since we are studying variable (flaring) sources, a reasonably
good representation of the observed spectrum can be obtained by
considering the particle distribution at the time $t_{\rm inj}$, i.e. at
the end of the injection, where in fact the emitted luminosity is
maximized.

Beside the geometrical difference, clearly the main (and
possibly crucial as far as the interpretation is concerned) difference
between the steady--state and the finite injection hypothesis is
related to the shape of the emitting particle distribution.
In fact in the latter scenario $\gamma_{\rm peak}$ does not have to be
associated with the minimum energy of the injected particles (as
was the case so far in our modeling). 
If the injection lasts for a finite timescale, only the higher energy
particles have the time to cool and therefore the particle
distribution can be described as a broken power--law with a steep part above
$\gamma_{\rm peak}$ and the original injection slope below.
In other words $\gamma_{\rm peak}$ results to be defined by the condition
$t_{\rm cool}(\gamma_{\rm peak}) = t_{\rm inj}$, while the minimum energy
$\gamma_{\rm min}$ can be much smaller than $\gamma_{\rm peak}$.

More precisely, let us define $\gamma_{\rm c}$ as the energy of those
electrons that can cool in the injection time $t_{\rm inj}$,
i.e. $t_{\rm cool}(\gamma_{\rm c})=t_{\rm inj}$.
At energies greater than $\gamma_{\rm c}$, particles radiatively cool, and
the distribution reaches a steady state in a time smaller than
$t_{\rm inj}$.
The particle distribution $N(\gamma)$ will be then described according
to the following prescriptions:
\begin{itemize}
\item 
If $\gamma_{\rm c}>\gamma_{\rm max}$, then no particles cool in the injection
time and we assume that $N(\gamma)\propto \gamma^{-s}$ above
$\gamma_{\rm min}$, and $N(\gamma)\propto \gamma^{-1}$ below.
\item
If $\gamma_{\rm min}<\gamma_{\rm c}<\gamma_{\rm max}$, we assume that $N(\gamma)
\propto \gamma^{-(s+1)}$ above $\gamma_{\rm c}$, $N(\gamma) \propto
\gamma^{-s}$ between $\gamma_{\rm min}$ and $\gamma_{\rm c}$ and
$N(\gamma)\propto \gamma^{-1}$ below $\gamma_{\rm min}$.
\item
If $\gamma_{\rm c} < \gamma_{\rm min}$ then 
$N(\gamma) \propto \gamma^{-(s+1)}$ above $\gamma_{\rm min}$; 
$N(\gamma) \propto \gamma^{-2}$ between $\gamma_{\rm c}$ and $\gamma_{\rm min}$ 
and $N(\gamma)\propto \gamma^{-1}$ below $\gamma_{\rm c}$.  
\item
If electron of all energies radiatively cool in the time $t_{\rm inj}$, then 
$N(\gamma) \propto \gamma^{-(s+1)}$ above $\gamma_{\rm min}$ and 
$N(\gamma) \propto \gamma^{-2}$ below.
\end{itemize}
We note here that even if the shape of $N(\gamma)$ is not formally
derived by solving the continuity equation, 
it does correspond to the shape expected from injecting a
broken power--law with a break at $\gamma_{\rm min}$, and slopes $\propto
\gamma^{-1}$ and $\propto \gamma^{-s}$ below and above.

With these prescriptions we then modeled all SEDs.  
In Table 3 we list
the input parameters used (plus the derived value of 
$\gamma_{\rm peak}$) for the spectra shown in Fig.~1a,b,c as dashed lines.  
% {\bf bisogna mettere anche i parametri per le sources di G98? 
% Non ricordo se ne abbiamo gia parlato} 
It is apparent that the accuracy of the representation of the SED is
comparable with that obtained in the stationary assumption, and does
not allow to discriminate between the two models.

In Fig.~3 we show $\gamma_{\rm peak}$ as a function of the comoving energy
density $U$ according to the application of this model to $all$
sources, the sources analyzed in this
paper plus those considered in G98 for consistency.
As expected, for powerful blazars (large $U$, small $\gamma_{\rm peak}$) we do not find significant
differences with respect to the results of the previous model (compare
the branches at high $U$ of Fig.~2 and Fig.~3), since for these 
sources $\gamma_{\rm c} \ll \gamma_{\rm min}$, and therefore 
$\gamma_{\rm peak}\sim \gamma_{\rm min}$ in both scenarios.
However, for small values of $U$, $\gamma_{\rm peak}$ coincides with
$\gamma_{\rm c} >\gamma_{\rm min}$ and this automatically ensures that 
$\gamma_{\rm peak}\propto U^{-1}$, except when $\gamma_{\rm c}$ results 
to be so large to exceed $\gamma_{\rm max}$ (in this case, of course,
$\gamma_{\rm peak}=\gamma_{\rm max}$).
In the finite injection model, in fact, the relation
$\gamma_{\rm c}\propto U^{-1}$ is built--in, and translates
into a $\gamma_{\rm peak} \propto  U^{-1}$ 
when $\gamma_{\rm c}>\gamma_{\rm min}$. 
What we have verified is that this model fits the SEDs
of all blazars, including the powerful ones.
For the latter (i.e. large $U$ and small $\gamma_{\rm peak}$) 
we find the same correlation as found using the steady state model.
We conclude that also this more consistent scenario
confirms the existence of the two branches.
These connect, as before, for values of $U\sim$ a few erg cm$^{-3}$
and $\gamma_{\rm peak}\sim 300$. 
This is a consequence of $\gamma_{\rm c}$ becoming smaller
than $\gamma_{\rm min}$ caused by the increased radiative
cooling in powerful blazars.

\section{Summary and conclusions}

By studying blazars of very high and very small observed power we
have confirmed that different flavors of blazars form a spectral
and power sequence, where the peak frequency position and the relative
intensity of the low and high energy spectral components decrease with
increasing source power.

This phenomenological behavior is plausibly the manifestation of a
physical trend.
By (necessarily) adopting an emission model for the production of the
radiation, it is then possible to infer the physical parameters
and look for the process(es) responsible for it.

In particular we considered the emission via synchrotron and inverse
Compton scattering by a homogeneous region containing a tangled
magnetic field and relativistic electrons. 
Already in a previous study (G98) this resulted in the finding  
of a clear relationship  
between the energy of the particles emitting at the peaks of the 
spectrum $\gamma_{\rm peak} m_e c^2$ and the total energy density. 
Here we extended the range of parameters by including  
sources with more extreme values of $\gamma_{\rm peak}$. 
The physical conditions found for low power BL Lacs
imply a radiative cooling timescale long compared with the 
source light crossing time, and led us to consider
the effects of a finite time $t_{\rm inj}$ for the injection of the 
relativistic particles.

The modeling of the blazar SED including a finite injection 
timescale (e.g. plausibly resembling what expected in 
the internal shock scenario) confirms 
the existence of a new branch of the correlation at high 
$\gamma_{\rm peak}$, with $\gamma_{\rm peak}\propto U^{-1}$.
As the injection of particles
above an energy $\gamma_{\rm min} m_{\rm e} c^2$ lasts for $t_{\rm inj}$ we 
obtain two behaviors:
in the fast cooling regime $\gamma_{\rm peak}\sim\gamma_{\rm min}$
and we re--obtain the result of G98,
while in the slow cooling regime $\gamma_{\rm peak}$ corresponds to
particles whose cooling time equals the injection time.
In this case $\gamma_{\rm peak}$ is always greater than $\gamma_{\rm min}$
and we obtain $\gamma_{\rm peak}\propto U^{-1}$.

While these results and their interpretation do not reveal the
acceleration mechanism itself, they are suggestive of the fact that
two processes maybe at work: a phase of (pre--)heating determining
$\gamma_{\rm min}$, and a phase of rapid acceleration leading to a 
non--thermal distribution.
If so the typical energy produced by the pre--heating 
(in the range $\gamma_{\rm min} \sim$ 30--10$^3$)
would correspond to the balancing between heating and cooling rates,
which would give the $\gamma_{\rm min}\propto U^{-0.5}$ dependence.
The second phase, reminiscent of  acceleration at shocks, would have to be
a fast (``instantaneous") acceleration of particles according to a
power--law distribution. 
The energetic particles would then cool (as in post--shock plasma) and
after a time $t_{\rm inj}$ the particles above $\gamma_{\rm c}$ would have 
cooled according to $\gamma_{\rm c}\propto U^{-1}$.
Although this interpretation might sound rather speculative, we
stress that the occurrence of these two phases is 
indeed needed in the context of particle acceleration by shocks.

We also note that, in the framework of the adopted finite time
injection model,
we cannot reproduce the SED of blazars by imposing that a single mechanism
is at work, namely by fixing $\gamma_{\rm min}$ to a constant 
value for all blazars and letting only $\gamma_{\rm c}$ to assume
the value appropriate for the particular cooling conditions.
In this case, in fact, we would be forced to assume
$\gamma_{\rm min}\sim$a few for all blazars (to properly fit the powerful
ones) with a particle distribution $N(\gamma)\propto \gamma^{-s}$
between $\gamma_{\rm min}$ and $\gamma_{\rm c}$.
As a result, we would overestimate the flux of our extreme BL Lacs
at frequencies below the synchrotron peak,
which instead require $\gamma_{\rm min} \gsim 10^3$. 

% {\bf Pre--heating, if energy is equally shared between electron and
% protons, makes electron highly relativistic.  Correspondingly, their
% Larmor radius becomes comparable with the Larmor radius of protons,
% and this favors cases in which the shock acceleration process does not
% deposit more energy into protons than into electrons. Dici di metterlo?}

An important point to be considered in interpreting the above findings
is that they refer to an average state of the source within the 
proposed blazar sequence.
However, individual flares have been observed to behave also
differently: a specific source can vary by a significant factor in the
direction opposite to the sequence trend, 
i.e.  both $\gamma_{\rm peak}$ and the observed
luminosity can increase at the same time (e.g. Mkn 501
during its 1997 active state).
The correlation between $\gamma_{\rm peak}$ and the
comoving energy density still holds even in these cases (see 
the ``tracks" shown in Fig. 2 and Fig. 3 for the TeV BL Lacs modeled  
in different states), but it is important to notice that 
in the case of the flaring state of Mkn 501 the particle distribution 
is found to be quite flat (i.e. $N(\gamma)\propto \gamma^{-(s+1)}$ 
with $s<2$), at variance with the typical values generally found 
(i.e. $s>2$, see Table 3).
This implies values of $\gamma_{\rm peak}$ close to $\gamma_{\rm max}$
irrespective of the amount of radiative cooling.
Physically, this suggests that when the source
undergoes major flares, the shock acceleration mechanism
becomes energetically dominant with respect to the pre--heating
process, leading to flat particle distributions and thus to flat spectra
in the entire synchrotron frequency range.

% Another limit (rather, a saturation)
% of the found correlation occurs since $\gamma_{\rm peak}$ cannot be
% greater than $\gamma_{\rm max}$, whose value cannot be determined with
% certainty.  

Let us conclude by asking whether there is any physical limit to the
value of $\gamma_{\rm peak}$.
BL Lac objects even more extreme than Mkn 501 could exist, with
their synchrotron peak frequency reaching the MeV band.   
The found trends suggest that the most extreme values of
$\gamma_{\rm peak}$ are possible only in sources intrinsically weak 
in the
radio and optical band which could have thus escaped so far from being
recognized as BL Lac objects in existing samples (see e.g. Ghisellini 1999b).
As in the radio and optical band the SED 
would presumably be dominated by the emission
from the host galaxy, these putative sources 
might resemble low power radio galaxies
(such as those of the B2 sample). Their
BL Lac--ness should appear evident in the TeV band and at hard X--ray
energies, although care should be taken not to confuse their emission with
the similarly hard X--ray spectra produced by e.g. low radiative
efficiency accretion model (where the radiation is via free free or
thermal Comptonization).

%{\bf precursor in blazaroni? }

\begin{acknowledgements}
We thank Marek Sikora for useful
discussions. AC thanks the Italian MIUR for financial support.
This research has made use of the NASA/IPAC Extragalactic Database (NED)
which is operated by the Jet Propulsion Laboratory, Caltech, under contract
with the National Aeronautics and Space Administration.
\end{acknowledgements}


\begin{thebibliography}{}

\bibitem[]{} Bersanelli et al. 1992, AJ, 104, 28 (Be92)
\bibitem[]{} Bohringer H., Voges W., Huchra J.P. et al., 2000, ApJS, 129, 435
\bibitem[]{} Brinkmann W., Siebert J., Feigelson E.D. et al., 
             1997, A\&AS, 323, 739 (Br97)
\bibitem[]{} Brinkmann W. \& Siebert, 1994, A\&A, 285, 812 (BS94)
\bibitem[]{} Catanese M., \& Weekes T.C., 1999, PASP, 111, 1193 
\bibitem[]{} Chiaberge M. \& Ghisellini G., 1999, MNRAS, 306, 551 
\bibitem[]{} Costamante L., Ghisellini G., Giommi P. et al., 2001, 
        A\&A, 371, 512  (Co01)
 \bibitem[]{} Fabian A.C., Celotti A., Iwasawa K., McMahon R.G.,
             Carilli C.L., Brandt W. N., Ghisellini G. \& Hook I.M., 
             2001a, 323, 373 (Fa01a)
\bibitem[]{} Fabian A.C., Celotti A., Iwasawa K. \& Ghisellini G., 2001b, 
             MNRAS, 324, 628 (Fa01b)
\bibitem[]{} Falomo R., Scarpa R. \& Bersanelli M., 1994, ApJS, 93, 125 (Fa94)
\bibitem[]{} Falomo R. \& Kotilainen J., 1999, A\&A, 352, 85 
% \bibitem[]{} Fegan 1996{\bf MANCA ???}
\bibitem[]{} Fichtel C.E., Bertsch D.L., Chiang J., et al. 
             1994, ApJS, 94, 551 (Fi94)
\bibitem[]{} Fossati G., Maraschi L. Celotti A., Comastri A. \&
             Ghisellini G., 1998, MNRAS, 299, 433 (F98)
% \bibitem[]{} Fossati G., Celotti A., Chiaberge M. et al. 2000, ApJ, 541, 166 
\bibitem[]{} Ghisellini G., Celotti A., Fossati G., Maraschi L. \& Comastri A.,
             1998, MNRAS, 301, 451 (G98)
\bibitem[]{} Ghisellini G., 1999a, 4th ASCA Symp., Astronomische Nachrichten. 
         Editors: H. Inoue, T.Ohashi \&  T.Takahashi, 320, p. 232  
\bibitem[]{} Ghisellini G., 1999b, in: TeV Astrophysics of Extragalactic 
             Objects, Astroparticle Physics, Eds. M. Catanese \& T. Weeks, 
             Vol. 11,  Issue 1--2, p. 11--18 
\bibitem[]{} Giommi P., Ansari S.G. \& Micol A., 1995, A\&AS, 109, 267 (GAM95)
\bibitem[]{} Giommi P. \& Padovani P., 1994, MNRAS, 268, L51
\bibitem[]{} Giommi P., Padovani P. \& Perlman E., 2000, MNRAS, 317, 743 (GI00)
\bibitem[]{} Horan D. VERITAS collaboration, 2000, HEAD meeting, No 23, 
             05.03 (Ho00)
\bibitem[]{} Kataoka J., Mattox J.R., Quinn J., et al., 1999, ApJ, 514, 138 
\bibitem[]{} Kino M., Takahara F. \& Kusunose M., 2001, ApJ, in press 
             (astro--ph/0107436)
\bibitem[]{} Lamer G., Brunner H., \& Staubert R., 1996, A\&A, 311, 384 (La96)
\bibitem[]{} Macomb J., Akerlof C.W., Aller H.D. et al. 1995, ApJ, 449, L99 
             (Ma95)
\bibitem[]{} Macomb J., Akerlof C.W., Aller H.D. et al. 1996, ApJ, 459,
             L111 (erratum) (Ma96)
\bibitem[]{} Mastichiadis A. \& Kirk J.G., 1997, A\&A, 320, 19
\bibitem[]{} Moran E.C. \& Helfand D.J., 1997, ApJ, 484, L95 (MH97)
\bibitem[]{} Padovani P. \& Giommi P., 1995, ApJ, 444, 567
\bibitem[]{} Padovani P. \& Urry C.M., 2001, Blazar Demographics and
Physics, ASP Conf Series, 227
\bibitem[]{} Perlman E.S, Stocke J.T., Schachter J.F., et al., 
             1996, ApJS, 104, 251 (Pe96)
\bibitem[]{} Pian E. \& Treves A., 1993, ApJ, 416, 130 (Pi93)
\bibitem[]{} Pian E., Vacanti G., Tagliaferri G. et al., 1998, ApJ, 491, L17  
\bibitem[]{} Piran T., 1999, Phys. Rep., 314, 575
\bibitem[]{} Sambruna R.M., Barr P., Maraschi L., Tagliaferri G. \&
             Treves A.,  1993, ApJ, 408, 452 (Sa93)
\bibitem[]{} Sambruna R.M., Barr P., Giommi P., Maraschi L., 
             Tagliaferri G. \& Treves A., 1994, ApJS, 95, 37 (Sa94)
\bibitem[]{} Sambruna R.M., George I.M., Madejski G., Urry C.M., Turner T.J., 
             Weaver K.A., Maraschi L. \&  Treves, A., 
             1997, ApJ, 483, 774 (Sa97)
\bibitem[]{} Schubnell M., 1997, Proceedings of the Fourth Compton Symposium, 
             Eds C.D. Dermer, M.S. Strickman \& J.D. Kurfess, AIP Conference 
             Proceedings 410, p. 1386 (astro--ph/9707047)
\bibitem[]{} Sikora M., Blazejowski M., Begelman M.C. \& Moderski R., 2001,
             ApJ 544, 1. 
\bibitem[]{} Sikora M. \& Madejski G., 2001, in: International 
             Symposium on High Energy Gamma-Ray Astronomy,  Heidelberg, Eds. 
             F. Aharonian and H. Voelk, 2001, AIP, in press (astro--ph/0101382)
\bibitem[]{} Stevens, J. \& Gear 1999,  MNRAS, 307, 403 (SG99)
\bibitem[]{} Stocke J.T., Morris S.L., Gioia I.M., Maccacaro T., Schild R.,
             Wolter A., Fleming T.A., Henry J.P.  1991, ApJS, 76, 813 
\bibitem[]{} Tagliaferri G., Ghisellini G., Giommi P. et al., 2001,
             A\&A, 368, 38 (Ta01)
\bibitem[]{} Tashiro M., Makishima K., Ohashi T., Inda--Koide M., Yamashita A., 
             Mihara T., Kohmura Y., 1995, PASJ, 47, 131 (Ta95)
\bibitem[]{} Tavecchio F., Maraschi L. \& Ghisellini G., 1998,
             ApJ, 509, 608 
\bibitem[]{} Wolter A., Tavecchio F., Caccianiga A., Ghisellini G. \&
             Tagliaferri G., 2000, A\&, 357, 429 (Wo00)
\bibitem[]{} Worral D.M. \& Wilkes B.J., 1990, ApJ, 360, 396 (WW90)
%      \bibitem[]{} Tagliaferri G., 2000, A\&A, 357, 429 
% 

\end{thebibliography}
\end{document}